\documentclass[aps,twocolumn,showpacs,preprintnumbers,amsmath,amssymb,floatfix]{revtex4}

\usepackage{graphicx}
\usepackage{longtable}
\usepackage{dcolumn}
\usepackage{bm}
\usepackage{color}
\usepackage[normalem]{ulem}
\usepackage[colorlinks=true, urlcolor=blue, linkcolor=blue, citecolor=blue]{hyperref}
\usepackage[all]{hypcap}
\newcommand{\beq}{\begin{equation}}
\newcommand{\eeq}{\end{equation}}
\newcommand{\bea}{\begin{eqnarray}}
\newcommand{\eea}{\end{eqnarray}}

\begin{document}

\title{Ab-initio calculation of the Hubbard $U$ and Hund exchange $J$ in local moment magnets:
The case of Mn-based full Heusler compounds}

\author{M. Tas$^{1}$}\email{murat.tas@gtu.edu.tr}
\author{E. \c{S}a\c{s}{\i}o\u{g}lu$^{2}$}\email{ersoy.sasioglu@physik.uni-halle.de}
\author{S. Bl\"{u}gel$^{3}$}
\author{I. Mertig$^{2}$}
\author{I. Galanakis$^{4}$}\email{galanakis@upatras.gr}
\affiliation{$^{1}$Department of Physics, Gebze Technical University, 41400 Kocaeli, Turkey \\
$^{2}$Institute of Physics, Martin Luther University Halle-Wittenberg, 06120 Halle (Saale),
Germany \\
$^{3}$Peter Gr\"{u}nberg Institut and Institute for Advanced Simulation, Forschungszentrum
Julich and JARA, 52425 Julich, Germany \\
$^{4}$Department of Materials Science, School of Natural Sciences, University of Patras,
GR-26504 Patra, Greece}

\date{\today}

\begin{abstract}
Mn-based full Heusler compounds possess well-defined local atomic Mn moments, and thus the
correlation effects between localized $d$ electrons are expected to play an important role in
determining the electronic and magnetic properties of these materials. Employing  ab-initio
calculations in conjunction with the constrained random-phase approximation (cRPA) method, we
calculate the strength of the effective on-site Coulomb interaction parameters (Hubbard $U$ and
Hund exchange $J$) in the case of X$_2$MnZ full Heusler compounds with X being one of Ni, Pd or
Cu, and Z being one of In, Sn, Sb or Te. We show that the Z element (or $sp$ element) in Heusler
compounds significantly reduces the strength of the Hubbard $U$ parameter for Mn 3$d$ electrons
compared to the elementary bulk Mn. On the contrary, the effect of the $sp$-atom on the strength of
the $U$ parameter of Ni, Cu or Pd valence $d$ electrons is not so substantial with respect to
the elementary bulk values. The $U$ values for all transition-metal atoms decrease with increasing
$sp$ electron number in the In-Sn-Sb-Te sequence. Our cRPA calculations reveal that despite their
well-defined local magnetic moments, the Mn-based full Heusler alloys fall into the category of
the weakly correlated materials.
\end{abstract}

\maketitle

\section{Introduction}\label{sec1}

Cu$_2$MnAl, the prototype Heusler compound, was synthesized more than a century ago
\cite{Heusler1903,Heusler1912}. Since this initial discovery, hundreds of compounds and alloys
crystallizing in similar lattices have been grown experimentally \cite{landolt,landolt2}. Due to
their large number and variety of the chemical elements, which can act as constituents, Heusler
compounds and alloys present a richness of physical phenomena \cite{landolt,landolt2,FelserRev2}.
Already in the sixties, full Heusler compounds with chemical formula X$_2$MnZ -where X is Ni, Pd
or Cu, and Z is In, Sn or Sb- were grown in the $L2_1$ lattice structure and were studied for
their magnetic properties \cite{Oxley1962,Webster1968,Webster1969}. These compounds are often
characterized as best metallic local moment magnets since the spin magnetic moment is concentrated
and localized at the Mn atoms, and the long range Mn-Mn exchange interactions determine the
magnetic behavior \cite{Kubler1983,Sasioglu2004,Sasioglu2008}. The interest on these Heusler
compounds has been intensified recently since it was shown that the off-stoichiometric Heusler
alloys containing Ni and Mn may present a martensitic phase transition which is accompanied by
strong magneto-caloric effects \cite{Buchelnikov1,Buchelnikov2,Buchelnikov3,Buchelnikov4}. This
class of materials is well-known in literature as ferromagnetic shape memory alloys
\cite{FSMA1,FSMA2,FSMA3,FSMA4,FSMA5}. Magnetic Heusler compounds and alloys are also promising
materials for the rapidly growing fields of magnetoelectronics and spintronics \cite{Elphick2021}.
Several aspects of the (Ni,Pd,Cu)$_2$Mn(In,Sn,Sb) Heusler compounds have been studied recently
including the nature of the exchange interactions \cite{Sasioglu2004,Sasioglu2008}, spin-waves
\cite{Galanakis2012,Fischer2020}, electronic properties \cite{Bose2010,Aguilera2019,Hamri2015},
the spin polarization \cite{Obaida2017}, and the martensitic phase transition
\cite{Pramanick2018,Wang2014}.

Although electronic band structure calculations based on the density functional theory (DFT) are
quite successful in the study of magnetic systems, the presence of 3$d$ and 4$d$ transition-metal
atoms in the local moment magnetic Heusler compounds means that the electronic correlations may
play an important role in the determination of their electronic and magnetic properties. For
example, it was shown in Ref. \cite{Fischer2020} for the compounds Ni$_2$MnSn and Pd$_2$MnSn that
the electronic correlations play a crucial role in correct determination of the spin-waves spectra.
Also as shown by Shourov and collaborators in the case of FeVSb, a semiconducting Heusler compound,
the inclusion of on-site correlations in the calculations leads to an enhancement of the effective electron mass by 40\% in agreement with their experimental findings \cite{Shourov}.

There are two common ways to include the correlations in the first-principles electronic structure
calculations. The first one is the so-called LDA+$U$ or GGA+$U$ (LDA acronym stands for ``local
density approximation", and GGA for ``generalised gradient approximation") scheme where an
effective on-site Coulomb repulsion term Hubbard $U$ and Hund exchange $J$ are used to account for
the correlation effects \cite{JepsenPRB2010,SolovyevReview}. This approach has been proven to be
accurate in the case of transition-metal oxides \cite{JepsenPRB2010,SolovyevReview}. But the
LDA+$U$ and GGA+$U$ schemes are not suitable for metallic systems and they cannot describe the
non-quasiparticle states in half metallic Heusler compounds \cite{Katsnelson}. A more elaborate
modern computational scheme, which has resulted from merging of the DFT and many-body Hamiltonian
methods, is the so-called LDA+DMFT where DMFT stands for dynamical mean field theory
\cite{MinarReview,DMFT}. The LDA+DMFT schemes have been proven to be successful in determining the
electronic properties of 3$d$ ferromagnets and several extensions including also the non-local
correlations have been proposed \cite{Katsnelson2}. Except LDA, also the GGA functional has been
used in conjunction with the DMFT method, and thus in the following when we use the term LDA+DMFT,
we also imply the GGA+DMFT scheme.

Determination of the Coulomb interaction parameters (Hubbard $U$ and Hund exchange $J$) from
experimental data is a very difficult task and only scarce data exist. Thus, it is highly
desirable to directly calculate these parameters, which are materials specific, from the first
principles \cite{Kotani,Solovyev,Schnell}. The earliest approach is the well-known constrained
local-density approximation (cLDA) \cite{cLDA1,cLDA2,cLDA3}. The cLDA is still in wide use even
though it is known to deliver unreasonably large $U$ and $J$ values especially for the late
transition-metal atoms due to the difficulties in compensating for the self-screening error of
localized electrons \cite{cLDA2}. Moreover, since the Hubbard $U$, within the cLDA approach, is
calculated in the framework of DFT, frequency dependence of the $U$ cannot be obtained. Contrary
to cLDA, the constrained random-phase approximation (cRPA) does not suffer from these deficiencies
and allows individual Coulomb matrix elements to be accessed -\textit{e.g.} on-site, off-site,
intra-orbital, inter-orbital, and exchange elements- including their frequency dependence
\cite{cRPA,cRPA_2,cRPA_3,Wehling,Miyake2008,cRPA_Sasioglu,Zhang_Brothers,Hunter,Kaltak}.
The cRPA method has been applied to a variety of material classes including elementary
transition-metal atoms \cite{Zhang_Brothers,JepsenPRB2010,cRPA_Sasioglu}, half-metallic Heusler
compounds \cite{Galanakis2013}, $f$-electron systems \cite{Nilsson}, double perovskites
\cite{Neroni2019}, oxides \cite{JepsenPRB2010} and transition-metal oxides, and perovskites
\cite{cRPA_2,Zhang_Brothers}. Here, we should note that the actual $U$ value used in the LDA+DMFT
calculations depends on the adopted low-energy model and may need to be considerably increased
with respect to its calculated value \cite{Casula}.

Motivated by the interest on the local moment magnets, we aim in the present work to determine
the strength of the Coulomb interaction parameters (Hubbard $U$ and Hund exchange $J$) using the
cRPA method for twelve Mn-based X$_2$MnZ (X=Ni, Cu, Pd, and Z=In, Sn, Sb, Te) full Heusler
compounds. This study provides a complete picture on the behavior of the Coulomb interaction
parameters in local moment Mn-based Heusler magnets upon variation of both X and Z atoms. Our
calculations have shown that the Z atoms play an essential role in determining the strength of
the effective Coulomb interaction between 3$d$ electrons of Mn atoms in these materials.
Specifically, strength of the $U$ for Mn-3$d$ electrons is substantially reduced with respect to
the corresponding value in bulk Mn. Moreover, the $U$ value decreases with increasing
$sp$-electron number in the In-Sn-Sb-Te sequence. The rest of the paper is organized as follows:
in Sec. \ref{sec2} we present details of our calculations, in Sec. \ref{sec3} we present and
discuss our results, and in Sec. \ref{sec4} we summarize and conclude.

\section{Computational Methodology}\label{sec2}

In this section we will briefly discuss the method for calculating the effective Coulomb
interaction parameters for the Mn-based Heusler compounds by providing references to previous
articles where the methodology is presented in detail.

The compounds with Z being In, Sn or Sb are well known to grow in the so-called $L2_1$ cubic
lattice of the Heusler compounds; the lattice is a f.c.c. with four atoms along the diagonal as
basis with Wyckoff positions: $\mathrm{X}=(0\: 0\: 0)$,
$\mathrm{Mn}=(\frac{1}{4}\: \frac{1}{4}\: \frac{1}{4})$,
$\mathrm{X}=(\frac{1}{2}\: \frac{1}{2}\: \frac{1}{2})$, and
$\mathrm{Z}=(\frac{3}{4}\: \frac{3}{4}\: \frac{3}{4})$ \cite{Oxley1962,Webster1968,Webster1969}.
The compounds with Te atom have not been grown yet experimentally, but we include them for
reasons of completeness. In Table \ref{table1} we include the lattice constants used in the
calculations which are the equilibrium lattice constants determined through the total energy
calculations. For all compounds the lattice constant exceeds 6 \AA\ and for the same Z atom it
increases in the sequence Ni-Cu-Pd.

\begin{figure*}
\begin{center}
\includegraphics[scale=0.29]{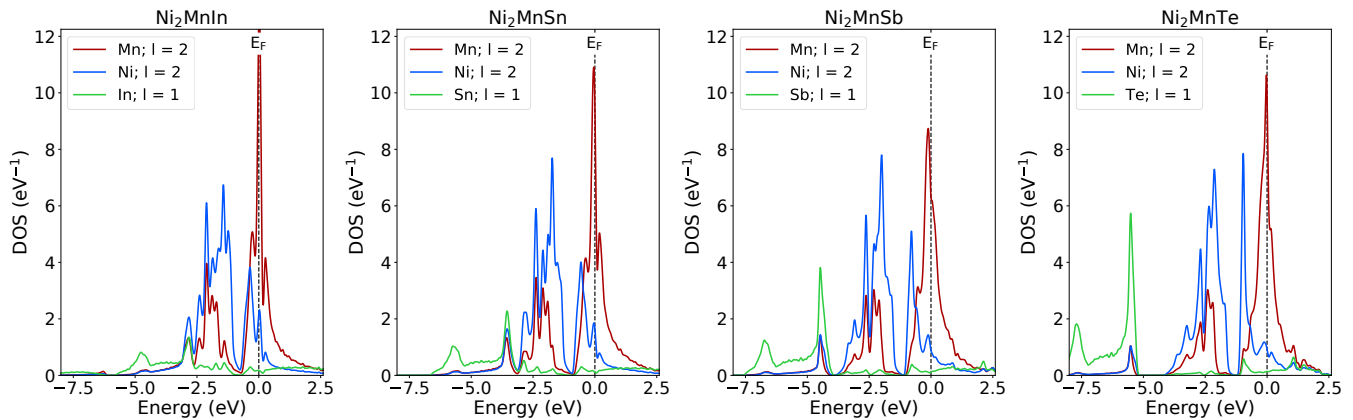}
\end{center}
\vspace*{-0.5cm}
\caption{(Colors online) Atom-resolved $l$-projected density of states (DOS) for non-magnetic
Ni$_2$MnIn, Ni$_2$MnSn, Ni$_2$MnSb, and Ni$_2$MnTe. For all compounds the Fermi level is set
to zero energy.}
\label{fig1}
\end{figure*}

The ground-state calculations for all studied full Heusler compounds were carried out using the
full-potential linearized augmented plane waves (FLAPW) method as implemented in the FLEUR code
\cite{Fleur} within the GGA of the exchange-correlation potential as parameterized by Perdew,
Burke and Ernzerhof (PBE) \cite{GGA}. Note that due to the metallic character of the compounds
under study, the GGA provides a more accurate description of the ground state properties with
respect to more complex hybrid functionals \cite{Meinert13}. For all calculations we used angular
momentum and plane-wave cutoff parameters of $l_{\textrm{max}}=10$ inside the spheres, and
$k_\textrm{max}=4.5 a_B^{-1}$ for the outside region. The DFT-PBE calculations were performed
using a $20 \times 20 \times 20$ \textbf{k}-point grid in the Brillouin zone. For compounds
including In atom, FLEUR uses local orbitals for semi-core $4d$ states.

\begin{figure}[!t]
\begin{center}
\includegraphics[scale=0.75]{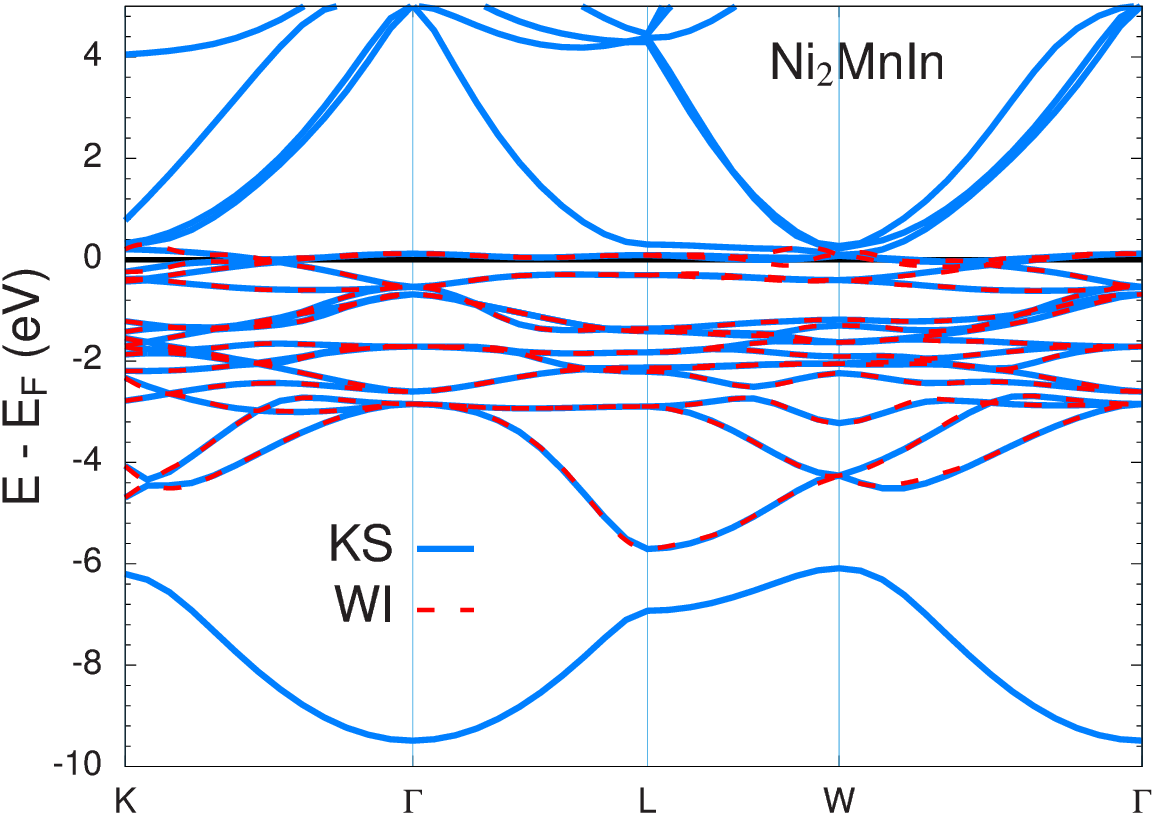}
\end{center}
\vspace*{-0.5cm}
\caption{(Colors online) Comparison of the DFT (KS) and the Wannier-interpolated (WI) band
structures of Ni$_2$MnIn along the high symmetry lines in the Brillouin zone.}
\label{fig2}
\end{figure}

The maximally localized Wannier functions (MLWFs) were constructed with the WANNIER90 code
\cite{Max_Wan,Wannier90,Mostofi,Marzari,Souza,Fleur_Wannier90}. The effective Coulomb potential
was calculated within the cRPA method \cite{cRPA,cRPA_2,cRPA_3} implemented in the SPEX code
\cite{Spex} (for further technical details see Ref. \cite{Sasioglu}). We used a
$8 \times 8 \times 8$ \textbf{k}-point grid, and 500 unoccupied bands in the cRPA calculations.
We also performed test calculations with a $6 \times 6 \times 6$ \textbf{k}-point grid; the
difference in the calculated parameters was of the order of 0.01 eV in most cases compared to
the more dense case. In full Heusler compounds, the Mn-3$d$ states are strongly hybridized with
the Ni-3$d$ (Cu-3$d$, Pd-4$d$) states as well as with the 5$p$ states of the Z atom as revealed
by the density of states (DOS) plots presented for the case of Ni-compounds in Fig. \ref{fig1}
(the bands corresponding to the Mn-3$d$ states scan the same energy window as the bands for the
Ni-3$d$ states while a less pronounced overlap exists also with the Z valence $p$ states).
Thus we construct the Wannier functions for the 3$d$ (4$d$) orbitals of the transition-metal
atoms using at least 19 bands. In some parts of the Brillouin zone a few more states are
included due to the mixing with other bands.

The chosen 15 Wannier orbitals well represent the correlated Hilbert space, for which in
Fig.\,\ref{fig2} we compare the Wannier-interpolated (WI) band structure with the DFT  one, denoted as KS (Kohn-Sham),
for the case of Ni$_2$MnIn. We see that the agreement for Ni$_2$MnIn is almost perfect. Figs
S1-S3 in the Supplementary Material present a similar comparison for the other three studied
compounds containing Ni \cite{Suppl}. In their case there are parts of the band structure where
the Wannier-interpolated band structure deviates from the KS one due to the $p$ admixture
discussed above. Moreover, in the fourth column of Table \ref{table1} we present the average
spread of the Wannier functions for all materials. The Wannier spreads provide qualitative
information on the localization of the Wannier functions, which will be discussed in the
following section. Note that as shown in Ref. \cite{Garrity}, the Wannier spreads do not
correlate with the accuracy of the Wannier interpolation scheme. Finally, we present, in the
Supplementary Material, the center of all Wannier functions and their respective spreads for
all compounds (see Tables S1-S12 in \cite{Suppl}).

\begin{table*}[t]
\caption{The lattice constants, $d$-orbitals, average spread of the Wannier functions $\Omega$,
bare (unscreened) $V$ and $J_b$, partially screened Hubbard $U$ and Hund exchange $J$,
$U_{\mathrm{eff}}$, and fully screened on-site Coulomb interaction parameters $\tilde{U}$ and
$\tilde{J}$ between the localized $d$-orbitals in twelve Heusler compounds. In parenthesis we
provide computed $U$ and $J$ values which are calculated by excluding 5$p$ screening channel
of the Z atom (see text for a detailed discussion).}
\begin{ruledtabular}
\begin{tabular}{lcllccccccc}
Compound   & a(\AA) & Orbital & $\Omega$ (\AA$^2$) & $V$ (eV) & $J_{\mathrm{b}}$ (eV) & $U$ (eV) &
$J$ (eV) & $U_{\mathrm{eff}}$ (eV) & $\tilde{U}$ (eV) & $\tilde{J}$ (eV)  \\ \hline
           &      &         &       &      &             &             &      &      &      &  \\
Ni$_2$MnIn & 6.08 & Ni-$3d$ & 1.88  & 23.01 & 1.16 & 3.60 (5.58) & 1.03 (1.06) & 2.57 & 0.75 & 0.73  \\
           &      & Mn-$3d$ & 7.17  & 18.97 & 0.96 & 2.60 (5.48) & 0.83 (0.88) & 1.77 & 0.13 & 0.28  \\
Ni$_2$MnSn & 6.02 & Ni-$3d$ & 2.12  & 22.66 & 1.14 & 3.43 (4.70) & 1.00 (1.02) & 2.43 & 0.88 & 0.76  \\
           &      & Mn-$3d$ & 7.72  & 18.51 & 0.93 & 2.57 (4.62) & 0.80 (0.85) & 1.77 & 0.14 & 0.29  \\
Ni$_2$MnSb & 6.00 & Ni-$3d$ & 1.84  & 22.23 & 1.11 & 3.07 (3.77) & 0.96 (0.97) & 2.11 & 1.06 & 0.78  \\
           &      & Mn-$3d$ & 8.70  & 17.85 & 0.89 & 2.33 (3.57) & 0.76 (0.79) & 1.57 & 0.16 & 0.31  \\
Ni$_2$MnTe & 6.07 & Ni-$3d$ & 1.72  & 21.81 & 1.08 & 2.47 (2.65) & 0.92 (0.92) & 1.55 & 1.02 & 0.76  \\
           &      & Mn-$3d$ & 8.24  & 16.82 & 0.82 & 1.62 (1.96) & 0.67 (0.69) & 0.95 & 0.12 & 0.24  \\
Pd$_2$MnIn & 6.37 & Pd-$3d$ & 2.41  & 15.70 & 0.88 & 3.00 (4.79) & 0.77 (0.82) & 2.23 & 1.26 & 0.68  \\
           &      & Mn-$3d$ & 7.74  & 19.42 & 0.99 & 2.34 (6.12) & 0.85 (0.94) & 1.49 & 0.11 & 0.25  \\
Pd$_2$MnSn & 6.38 & Pd-$3d$ & 2.03  & 15.65 & 0.87 & 2.94 (4.29) & 0.76 (0.80) & 2.18 & 1.36 & 0.69  \\
           &      & Mn-$3d$ & 8.90  & 19.04 & 0.96 & 2.36 (5.28) & 0.82 (0.90) & 1.54 & 0.11 & 0.25  \\
Pd$_2$MnSb & 6.42 & Pd-$3d$ & 2.01  & 15.54 & 0.86 & 2.66 (3.32) & 0.74 (0.75) & 1.92 & 1.44 & 0.68  \\
           &      & Mn-$3d$ & 10.78 & 18.44 & 0.92 & 2.03 (3.80) & 0.78 (0.83) & 1.25 & 0.11 & 0.25  \\
Pd$_2$MnTe & 6.35 & Pd-$3d$ & 2.19  & 15.42 & 0.85 & 2.47 (2.66) & 0.73 (0.73) & 1.74 & 1.44 & 0.68  \\
           &      & Mn-$3d$ & 13.59 & 17.55 & 0.87 & 1.69 (2.20) & 0.71 (0.74) & 0.98 & 0.11 & 0.22  \\
Cu$_2$MnIn & 6.19 & Cu-$3d$ & 1.07  & 24.95 & 1.27 & 4.75 (6.81) & 1.16 (1.23) & 3.59 & 2.67 & 1.08  \\
           &      & Mn-$3d$ & 9.82  & 17.95 & 0.89 & 1.77 (4.23) & 0.73 (0.91) & 1.04 & 0.15 & 0.29  \\
Cu$_2$MnSn & 6.20 & Cu-$3d$ & 0.99  & 24.82 & 1.26 & 4.38 (5.63) & 1.14 (1.19) & 3.24 & 2.71 & 1.08  \\
           &      & Mn-$3d$ & 11.24 & 17.07 & 0.84 & 1.54 (3.01) & 0.67 (0.78) & 0.87 & 0.14 & 0.27  \\
Cu$_2$MnSb & 6.10 & Cu-$3d$ & 1.18  & 24.64 & 1.25 & 3.89 (4.82) & 1.12 (1.17) & 2.77 & 2.70 & 1.07  \\
           &      & Mn-$3d$ & 12.23 & 15.73 & 0.75 & 1.10 (1.83) & 0.58 (0.69) & 0.52 & 0.14 & 0.26  \\
Cu$_2$MnTe & 6.27 & Cu-$3d$ & 1.09  & 24.51 & 1.24 & 3.48 (4.34) & 1.10 (1.15) & 2.38 & 2.54 & 1.06  \\
           &      & Mn-$3d$ & 10.94 & 14.05 & 0.65 & 0.55 (1.23) & 0.44 (0.62) & 0.11 & 0.09 & 0.18  \\
\end{tabular}
\label{table1}
\end{ruledtabular}
\end{table*}

In the cRPA approach, the full polarization matrix $P$ is divided into two parts: $P=P_d + P_r$,
where $P_d$ includes only \textit{d}-\textit{d} transitions and $P_r$ is the remainder. Then, the
frequency-dependent effective Coulomb interaction is given schematically by the matrix equation
$U(\omega)=[1 - vP_r(\omega)]^{-1}v$, where $v$ is the bare Coulomb interaction and $U(\omega)$
is related to the fully screened interaction by
$\tilde{U}(\omega) = [1 - U(\omega)P_d(\omega)]^{-1}U(\omega)$.

We consider matrix elements of $U$ in the MLWF basis
\begin{equation}
\begin{gathered}
U_{in_{1},jn_{3},in_{2},jn_{4}}(\omega) = \\
\int \int d\boldsymbol{r}d\boldsymbol{r}^\prime w_{in_1}^*(\boldsymbol{r})
w_{jn_3}^*(\boldsymbol{r}^\prime) U(\boldsymbol{r},\boldsymbol{r}^\prime,\omega)
w_{jn_{4}}(\boldsymbol{r}^\prime) w_{in_2}(\boldsymbol{r})\,.
\end{gathered}
\label{hubudef31}
\end{equation}
\newline

In the static limit ($\omega \longrightarrow 0$) the so-called Slater parametrization of the
Coulomb matrix is given by
\begin{equation}
U = \frac{1}{L^{2}} \sum_{m,n} U_{mn;mn} = F^{0} \,,
\label{Slater_U}
\end{equation}
\begin{eqnarray}
J &=& U - \frac{1}{L(L-1)}\sum_{m \neq n}[U_{mn;mn} - U_{mn;nm}] \\
&=& (F^{2} + F^{4})/14 \,, \nonumber
\label{Slater_J}
\end{eqnarray}
where $L$ is the number of localized orbitals, i.e., five for $d$ orbitals, $F^{0}$, $F^{2}$ and
$F^{4}$ are the Slater integrals. Similar to $U$ and $J$, we can also define so-called fully
screened $\tilde{U}$ and $\tilde{J}$ parameters as well as unscreened (or bare) $V$ and $J_b$
parameters. Although the fully screened Coulomb interaction parameters are not used in model
Hamiltonians, they provide an idea about the correlation strength of considered electrons.

\section{Results and Discussions}\label{sec3}

As mentioned in the preceding section, due to the strong hybridization of Mn-3$d$ states with
Ni-3$d$ (Pd-4$d$, Cu-3$d$) as well as the 5$p$ states of the Z atom, all twelve Heusler compounds
can be described by an effective fifteen-orbital low-energy model, five 3$d$ orbitals stemming
from the Mn atom, and ten 3$d$ (4$d$) orbitals stemming from the Ni, Cu (Pd) atoms. Note that in
the model Hamiltonian description of the Heusler compounds, the noninteracting one-body part of
the effective model is defined for a non-magnetic state, and thus calculation of the effective
Coulomb interaction parameters should be based on the same state. In the following, we will
discuss the effective Coulomb interaction parameters calculated within the cRPA method considering
the Slater parametrization of the Coulomb matrix for the non-magnetic state.

In Table \ref{table1} we present all computed results for all twelve compounds studied. The
calculations are material specific, and for each compound we provide computed values of the
bare (unscreened) Coulomb interaction parameters $V$, $J_{\mathrm{b}}$, partially screened
(Hubbard $U$, Hund exchange $J$) as well as fully screened interaction parameters for both X and
Mn atoms. In the non-magnetic case, the $V$ parameters provide information on extension (or
spread $\Omega$) of the Wannier functions, while as for the fully screened Coulomb interaction
parameters of the Mn atom one should not attribute a physical meaning to them due to the strong
change of electronic structure, i.e., sharp Mn-3$d$ peaks around the Fermi level (see Fig.
\ref{fig1}), which gives rise to a strong screening. In the ferromagnetic case of all these
compounds, one can obtain larger fully screened Coulomb interaction parameters for the Mn atom
due to the large exchange splitting of the Mn-3$d$ states. On the other hand, for the X atoms
we expect similar parameters in the magnetic and non-magnetic cases since the exchange splitting
is negligibly small for these atoms, especially in the case of Cu-based compounds. In the
following the fully screened Coulomb interaction parameters will not be discussed as they are
presented purely for the sake of completeness.

For $3d$ elements, the bare $V$ Coulomb interaction increases with increasing $d$-electron
number, in agreement with previous calculations. In the case of In-based compounds, $V$ varies
from 19 eV for the Mn atom to 25 eV for the Cu atoms and this behavior can be attributed to the
localization of the Wannier functions with increasing nuclear charge. An increase in the nuclear
charge causes the 3$d$ wave functions to contract, which gives rise to the observed trend for
$V$ and $J_b$. As we move within the same column of the periodic table from $3d$ to $4d$
elements, i.e., from Ni to Pd, the $V$ value decreases due to the more delocalized character of
the $4d$ Wannier functions of the Pd atom. For all compounds the Wannier functions are slightly
delocalized as we move from In to Te, which is reflected in the calculated spreads and bare
Coulomb interaction parameters presented in Table \ref{table1}. For instance, in the case of
Ni-based compounds the $V$ for Ni-$3d$ electrons decreases from 23 eV to 21.8 eV, while for the
Mn-$3d$ electrons this reduction is slightly larger, from 19 eV to 16.8 eV.

\begin{figure*}[!t]
\begin{center}
\includegraphics[scale=0.35]{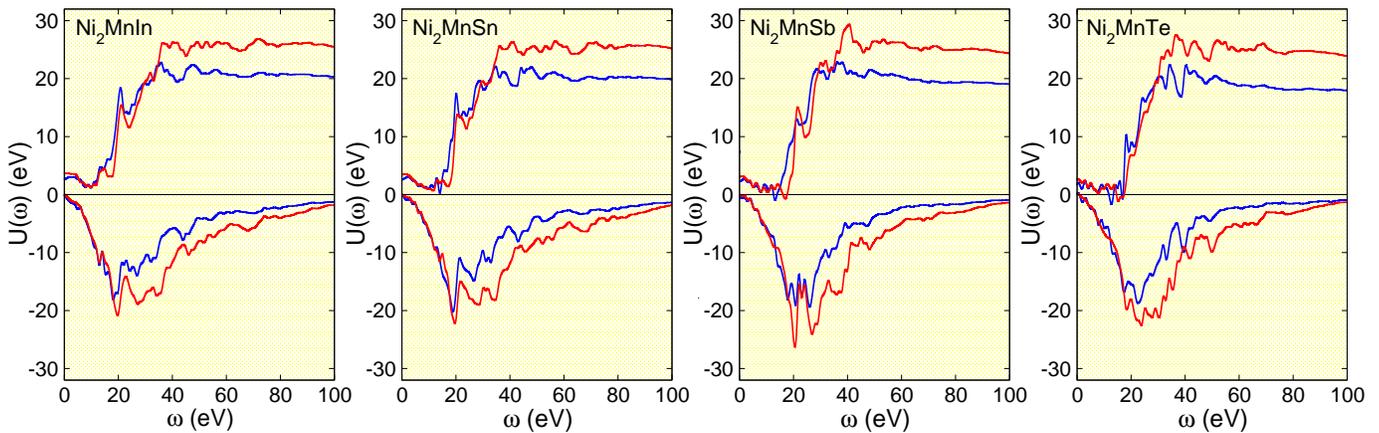}
\end{center}
\vspace*{-0.5cm}
\caption{(Colors online) Real (positive values) and imaginary (negative values) part of the
calculated on-site Coulomb interaction $U$ as a function of the frequency $\omega$ for the
Ni (red) and Mn (blue) atoms in the non-magnetic Ni$_2$MnIn, Ni$_2$MnSn, Ni$_2$MnSb, and
Ni$_2$MnTe compounds.}
\label{fig3}
\end{figure*}

We now would like to discuss the calculated Hubbard $U$ parameters. Obtained $U$ parameters for
X atoms in the In- and Sn-based compounds are more or less comparable to the corresponding
values in the elementary transition metals presented in
Refs. \cite{Zhang_Brothers,JepsenPRB2010,cRPA_Sasioglu}. With increasing $sp$-electron number,
i.e., along the In-Sn-Sb-Te sequence, the $U$ values for the X atoms decrease substantially,
especially in the case of Cu- and Ni-based compounds. As discussed above, this reduction
partially stems from the delocalization of the Wannier functions, but the main contribution
comes from the complex screening effects. Screening increases with increasing $sp$-electron
number, and thus results in smaller Coulomb interaction parameters. The same discussion holds
also for the effective Coulomb interaction parameter of Mn-$3d$ electrons. However the computed
$U$ values are significantly smaller than the corresponding value in elementary Mn atom, which
can be attributed to the efficient screening of $sp$-elements (Z atoms) in Heusler compounds.

Finally, we should discuss relative values of the on-site Coulomb interactions $U$ and of the
width $W$ of the $d$-bands. The ratio $U/W$ determines whether the material is a weakly or a
strongly correlated system. The situation is similar to the elementary 3$d$ transition metals
\cite{cRPA_Sasioglu} and the half-metallic Heusler compounds \cite{Galanakis2013}. Occupied
$d$-states, as shown in Fig. \ref{fig1}, scan an energy window $W$ of about 4-6\,eV width
depending on the specific Ni-based material. The $U$ values for the Ni and Mn atoms in these
compounds (presented in Table \ref{table1}) are smaller than $W$, which means that the $U/W$ ratio is
smaller than one and these materials are characterized as weakly correlated materials similar
to the elementary 3$d$ transition metals \cite{cRPA_Sasioglu} and the half-metallic Heusler
compounds \cite{Galanakis2013}. The same arguments stand also for the Pd- and Cu-based
compounds under study. Note that similarly to the studies in Refs. \cite{cRPA_Sasioglu} and
\cite{Galanakis2013}, the $U$ values do not differ significantly between the $e_g$ and $t_{2g}$
states; in Section I of the Supplementary Material \cite{Suppl} we present, as an example, the
$U_{mn;mn}$ values used in Eq.\,\ref{Slater_U} for the 3$d$ orbitals of of Ni and Mn atoms in
the Ni$_2$MnIn compound.

To reveal the contribution of the $sp$-element (Z atom) to screening of the effective Coulomb
interaction parameters, we present, in Table \ref{table1} inside the parenthesis, the $U$
values calculated with excluding the Z atom $5p$ screening channel, i.e., in the computation of
the polarization function (see Ref. \cite{cRPA_Sasioglu} for technical details) in addition to
the exclusion of the $3d \rightarrow 3d$ (or $3d \rightarrow 3d$ and $3d \rightarrow 4d$ in
Pd-based compounds) transitions we exclude also the $3d \rightarrow 5p$ ($3d \rightarrow 5p$
and $4d \rightarrow 5p$) transitions. As is seen, the Z atom $5p$ screening channel provides a
significant contribution to the strength of the Hubbard $U$ parameter in Heusler compounds. The
obtained $U$ values with and without the $5p$ screening channel differ more than by a factor of
two in some cases, especially in the case of In- and Sn-based compounds. In these two cases the
$5p$ channel is less than half filled and provides a substantial contribution to the screening
process. In the case of Te-based compounds, the $5p$ channel is more than half filled and thus
its contribution is reduced significantly.

Up to now we have discussed the strength of the effective Coulomb interaction parameter $U$ in
Heusler compounds. Unlike the $U$ parameter, the Hund exchange $J$ is much less screened as the
screening of the exchange interaction is monopole-like in contrast to dipole-like screening of
the $U$ parameter (see Ref. \cite{Miyake2008} for a detailed discussion). Thus, value of the
Hund exchange $J$ is close to the corresponding unscreened atomic value (see Table \ref{table1}).
Note that our computed $J$ values are larger than the ones presented in Refs.
\cite{cRPA_Sasioglu,Galanakis2013} for the elementary transition metals and half metallic
Heusler compounds. This difference is due to the different parametrization of the Coulomb matrix
since in those two cited papers the Hubbard-Kanamori parametrization is employed instead of the
Slater parametrization in the present work.

In Table \ref{table1} we also present the $U_{\mathrm{eff}}=U-J$, which is the so-called
effective Hubbard $U$. $U_{\mathrm{eff}}$ is used in Dudarev's approach which is a simplified
implementation of the DFT+$U$ method \cite{Dudarev,Dudarev2}. This approach, in conjunction
with GGA, was employed in Ref. \cite{Fischer2020} where the spin-wave spectra of Ni$_2$MnSn and
Pd$_2$MnSn have been calculated. There, an arbitrary $U_{\mathrm{eff}}$ value of 1.5 eV for the
Mn-3$d$ orbitals was considered. This value is very close to our calculated effective Hubbard
$U_{\mathrm{eff}}$ values for the Mn-3$d$ states shown in Table \ref{table1} (1.54 eV for
Pd$_2$MnSn and 1.77 eV for Ni$_2$MnSn). It was found that GGA+$U$ was a considerable improvement
over usual GGA calculations in reproducing the experimental spectra with accuracy, stressing the
importance of accurate determination of the Coulomb parameters.

\begin{table}[b]
\caption{Off-site (inter-sublattice) effective Coulomb interaction parameters in eV. X stands
for Ni, Pd or Cu depending on the chemical type of the compound. We should note that Mn-X are
the nearest neighbors while Mn-Mn and X-X are the next-nearest neighbors.}
\begin{ruledtabular}
\begin{tabular}{lccc}
X$_2$MnZ & $U_{\mathrm{Mn-X}}$(eV) & $U_{\mathrm{X-X}}$(eV) & $U_{\mathrm{Mn-Mn}}$(eV)  \\
\hline
Ni$_2$MnIn     & 0.47 & 0.38 & 0.19  \\
Ni$_2$MnSn     & 0.49 & 0.39 & 0.20  \\
Ni$_2$MnSb     & 0.44 & 0.35 & 0.18  \\
Ni$_2$MnTe     & 0.24 & 0.19 & 0.08  \\
Pd$_2$MnIn     & 0.36 & 0.31 & 0.13  \\
Pd$_2$MnSn     & 0.40 & 0.33 & 0.15  \\
Pd$_2$MnSb     & 0.29 & 0.23 & 0.09  \\
Pd$_2$MnTe     & 0.25 & 0.20 & 0.08  \\
Cu$_2$MnIn     & 0.39 & 0.31 & 0.16  \\
Cu$_2$MnSn     & 0.32 & 0.25 & 0.11  \\
Cu$_2$MnSb     & 0.20 & 0.14 & 0.07  \\
Cu$_2$MnTe     & 0.09 & 0.07 & 0.03  \\
\end{tabular}
\label{table2}
\end{ruledtabular}
\end{table}

In Table \ref{table2} we again present the Coulomb interaction parameters but now the off-sites
ones between $d$ orbitals of neighboring atoms are included in the table. These values, as
expected, are one order of magnitude smaller than the on-site Coulomb interaction parameters
discussed just above. The exact values depend not only on the chemical elements themselves but
also on the distance between neighboring atoms, and localization of the Wannier functions.
The Mn-X atoms are the nearest neighbors while the Mn-Mn and X-X pair of atoms are the next
nearest neighbors, and this explains the larger values in the first column. As we move from
one compound to another we cannot identify a very clear trend since variation of the off-site
$U$ value seems to depend on both the lattice constant and chemical elements. It is considerably
smaller when the Z atom is Te. Although one may conclude that the off-site Coulomb repulsion
terms can be neglected when the Hubbard-type model Hamiltonians are used to describe electronic
band structure, such a statement is an oversimplification since its validity depends on the
studied properties. The importance of the off-site Coulomb interactions for certain properties
(like, for example, the ones related to charge transfer) is enhanced due to the connectivity
of the lattice and the long-range $1/r$-tail of the partially screened Coulomb interaction
present in the cRPA treatment \cite{Schuler,Hansmann,Seth}, and these parameters might be
necessary for an accurate modeling of these materials.

Finally we would like to discuss the frequency dependence of the effective Coulomb interaction
parameter $U$ by considering the Ni-based compounds. In Fig. \ref{fig3} we plot both the real
(positive values) and imaginary (negative values) parts of the Coulomb interaction parameter
$U$ as a function of the frequency $\omega$. The $U$ values presented in Table \ref{table1} are
the static limit of the real part when the frequency tends to zero. For the Hubbard model to be
accurate, the $U$ values near the zero frequency should be quite stable as was the case in the
half-metallic Heusler compounds \cite{Galanakis2013}. This is not the case in the compounds
under study. With the exception of Ni$_2$MnIn, as we move away from the zero frequency, $U$
vanishes at the plasmon frequency and then shows an abrupt increase reaching the bare $V$ value
which remains almost constant for large $\omega$ values. Exactly at the plasmon frequency the
imaginary part of $U$ exhibits the first large peak and its value is around 20-25 eV for all
four Ni-based compounds in Fig. \ref{fig3}. Thus we expect that the static Hubbard models like
the LDA+$U$ and GGA+$U$ methods might not be so accurate in describing their electronic and
magnetic structure of these compounds. We should also mention that we have also investigated
behavior of the Hund exchange $J$ parameter as a function of $\omega$. The $J$ is not sensitive
to $\omega$, and its value remains almost constant for all considered frequency values
(it shows in all cases a very small increase of less than 0.1 eV at the plasmon frequency and
then remains constant).

\section{Conclusions}\label{sec4}

Employing the cRPA method within the FLAPW framework we have calculated the strength of the
on-site Coulomb interaction parameters (Hubbard $U$ and Hund exchange $J$) between the
localized $d$-electrons in X$_2$MnZ (X$=$Ni, Pd, Cu; Z$=$In, Sn, Sb, Te) Heusler compounds,
which are known to be best local moment magnets. Our calculations have shown that due to the
presence of the Z element (or $sp$ element) in Heusler compounds, the strength of the Hubbard
$U$ parameter for the Mn 3$d$ electrons is significantly reduced compared to the elementary
bulk Mn. In the case of the $d$ electrons of Ni, Cu and Pd, the strength difference of the $U$
parameter between the studied Heusler compounds and the elementary bulk Ni, Cu or Pd is not so
substantial. Moreover, the $U$ values for the transition metal valence $d$ electrons decrease
with increasing $sp$ electron number in the In-Sn-Sb-Te sequence. The calculated off-site
Coulomb parameters are one order of magnitude smaller than the on-site ones. Frequency
dependent calculations of the $U$ parameter reveal that the static limit might not be a good
approximation for these compounds, with the exception of the ones containing In.

Our cRPA calculations reveal that despite their well-defined local magnetic moments the
Mn-based full Heusler alloys fall into the category of the weakly correlated materials.
Knowledge of the Coulomb interaction parameters plays an important role in the construction of
model Hamiltonians aiming to study the correlation effects in electronic structure of ternary
magnetic compounds. We expect that our study will enhance the interest in Heusler
compounds which are local moment magnets.

\begin{acknowledgements}

E.\c{S}. and I.M. acknowledge support from Sonder-forschungsbereich TRR 227 of the Deutsche
Forschungs-gemeinschaft (DFG) and funding provided by the European Union (EFRE), Grant No.
ZS/2016/06/79307. Authors thank Christoph Friedrich for fruitful and helpful discussions.

\end{acknowledgements}

\end{document}